




\magnification=\magstep1
\hsize = 31.5pc
\vsize = 45pc

\baselineskip=15.7pt
\tolerance 2000
\parskip=6pt

\font\grand=cmbx10 at 14.4truept
\font\grand=cmbx10 at 14.4truept
\font\ninebf=cmbx9
\font\ninerm=cmr9

\def\d{\delta}
\def\p{\phi}

\def\pa{\partial}

\def\tr{{\rm tr}}
\def\ln{\,{\rm ln}\,}
\def\mr{M_{\rm red}}
\def\mc{M_{\rm c}}
\def\R{{\bf R}}

{
\baselineskip=12pt
\null
\leftskip=10.8cm{\noindent
INS-Rep.-1048\hfill\break
August 1994
}

\vskip 8mm
\vfill
}

{
\baselineskip=12pt
\centerline
{{\grand Global Aspects of the WZNW Reduction
to Toda Theories}\footnote*{\rm
Paper based on a talk given by I. Tsutsui at the Workshop on
{\it Quantum Field Theory, Integrable Models and Beyond}
of the Yukawa Institute for Theoretical Physics, 14 -- 17
February 1994.}
}

\vskip 7mm
\centerline{I. Tsutsui}
\vskip 3mm
\centerline{\it Institute for Nuclear Study}
\centerline{\it University of Tokyo}
\centerline{\it Midori-cho, Tanashi-shi, Tokyo 188}
\centerline{\it Japan}
\vskip 5mm
\centerline{and}
\vskip 5mm
\centerline{L. Feh\'er}
\vskip 3mm
\centerline{\it Laboratoire de Physique Theorique}
\centerline{\it ENS de Lyon}
\centerline{\it 46, all\'ee d'Italie,  F-69364 Lyon Cedex 07}
\centerline{\it France}

\vskip 1cm

{\parindent=25pt
\narrower\smallskip\noindent
\ninerm
{\ninebf Abstract.}\quad
It is well-known that the Toda Theories
can be obtained by reduction from
the Wess-Zumino-Novikov-Witten (WZNW)
model, but it is less known that
this WZNW $\rightarrow$ Toda reduction
is \lq incomplete'.  The reason
for this incompleteness
being that the Gauss
decomposition used to define the Toda fields
from the WZNW field is valid locally but not
globally over the WZNW group manifold,
which implies that actually the reduced system is
not just the Toda theory but has much richer structures.
In this note we furnish a framework
which allows us to study the reduced system globally,
and thereby present some preliminary results
on the global aspects.
For simplicity, we analyze primarily
0 $+$ 1 dimensional toy models
for $G = SL(n, {\bf R})$,
but we also discuss
the 1 $+$ 1 dimensional
model for
$G = SL(2, {\bf R})$ which corresponds to
the WZNW $\rightarrow$ Liouville reduction.
\smallskip
}

}

\vfill\eject

\noindent
{\bf 1. Introduction}
\medskip

In recent years the subject of integrable
models in 1 $+$ 1 dimensions, especially
conformally invariant ones, has been
attracting considerable
attention.  Among them are the standard
Toda theories governed by the Lagrangian
$$
{\cal L}_{\rm Toda}(\varphi )={\kappa \over 2}
\biggl[ \sum_{i,j=1}^l{1\over {2\vert \alpha_i\vert^2}}K_{ij}
\partial_\mu\varphi^i\partial^\mu \varphi^j
-\sum_{i=1}^l m_i^2\, \exp\Bigl(
{1\over 2}\sum_{j=1}^lK_{ij}\varphi^j\Bigr) \biggr] \ ,
\eqno(1.1)
$$
where
$\kappa$ is a coupling constant,
$K_{ij}$ is the Cartan matrix and
the $\alpha_i$ are the simple roots of the simple Lie algebra
${\cal G}$ of rank $l$.
These Toda theories
have been studied intensively over
the past several years,
with particular reference to
an application to two dimensional gravity,
since the Liouville theory emerges when the
underlying group $G$, for which ${\cal G} = {\rm Lie}(G)$,
is $SL(2, {\bf R})$.
One of the salient features
of the Toda theories is that they possess
as symmetry algebras so-called ${\cal W}$-algebras [1],
which are a polynomial extension
of the chiral Virasoro algebra.  It has been by now
well recognized [2] that both
the origin of the ${\cal W}$-algebras
and the integrability of the Toda theories
can be nicely understood by reducing the Toda theories from the
Wess-Zumino-Novikov-Witten (WZNW) model [3]\footnote*{
The space-time conventions are:
$\eta_{00}=-\eta_{11}=1$, $x^{\pm}={1\over 2}(x^0\pm x^1)$
and $\partial_\pm =\partial_0 \pm\partial_1$.
The WZNW field $g$ is periodic in $x^1$ with period $2\pi$.},
$$
S_{\rm WZ}(g)={\kappa \over 2}\int\,d^2x\,{\eta}^{\mu\nu}\, \tr
\,(g^{-1}\partial_\mu g)(g^{-1}\partial_\nu g)
-{\kappa \over 3}\int_{B_3}\, \tr\, (g^{-1}dg)^3 \ .
\eqno(1.2)
$$
The reduction is performed in the Hamiltonian
formalism by imposing a certain set
of first class constraints in the WZNW model, where
the connection between the
WZNW field $g \in G$
in (1.2) and the Toda fields $\varphi^i$
in (1.1) arises from
the Gauss decomposition,
$$
g=g_+\cdot g_0\cdot g_-\,.
\eqno(1.3)
$$
Here $g_{0,\pm }$
are from the subgroups $e^{{\cal G}_{0,\pm }}$ of $G$
where ${\cal G}_0$ is the Cartan subalgebra and
${\cal G}_{\pm}$ are the subalgebras consisting of
elements associated to positive or negative
roots --- hence the corresponding
decomposition in the algebra being
${\cal G}={\cal G}_+ + {\cal G}_0 + {\cal G}_-$.
Then the Toda fields are given
by the middle piece of the Gauss
decomposition,
$$
g_0 = \exp
    \Bigl( {1\over 2}\sum_{i=1}^l\varphi^i H_i \Bigr)\ ,
\eqno(1.4)
$$
where $H_i\in {\cal G}_0$ are the Cartan generators
associated to the simple roots $\alpha_i$.

However, the above picture of the
WZNW $\rightarrow$ Toda reduction
is not quite complete because the
Gauss decomposition (1.3) is valid only locally in
the neighbourhood of the identity $g = 1$ but not globally
over the group manifold $G$.
(This incompleteness has been noticed already in the
early work on the WZNW $\rightarrow$ Liouville
reduction [4].)
This suggests that, in general, the reduced theory
may be thought of as a system consisting
of several subsystems defined
on each of the patches introduced to cover the entire
group manifold, and that the Toda
theory is merely the subsystem given
on the Gauss decomposable
patch where (1.3) is valid.

Clearly, for understanding
what the WZNW reduction
really brings about we need to
know (i) what is a possible
general framework for a global
description of the
reduced system, and
(ii) what are the
physical implications of \lq being global\rq, rather
than just \lq being local' considering the Toda
theory only.
The purpose of this paper is to set out
an investigation toward these desiderata.
In order to elucidate the essence as well as to ease
the problem,
we simplify the situation by considering
primarily the toy models given by the same
reduction in 0 $+$ 1 dimension, that is, we just neglect
the spatial dimension in the usual
WZNW $\rightarrow$ Toda reduction.  By doing so, the
WZNW model becomes a system of a particle moving freely
on the group manifold $G$, and the reduction renders
the reduced configuration space essentially
flat with a diminished dimension,
giving rise to a Toda type potential.
For brevity, we call those reduced toy models
\lq 0 $+$ 1 dimensional Toda theories', and
set $\kappa = 1$ throughout.

The plan of the present paper is the following:
We first illustrate in sect.2 an interesting
physical effect caused by
being global in a simple setup,
where we take up a 0 $+$ 1 dimensional
Toda theory
for $G = SL(2,\R)$, {\it i.e.}, the Liouville toy model.
We shall observe that a locally catastrophic motion of a particle
can be interpreted as an aspect of a globally
stable motion, an oscillation.
Then in sect.3 we provide
a framework for $G = SL(n, \R)$
which enables us to discuss the reduction globally.
This will be done in two ways --- first by furnishing
a decomposition (Bruhat decomposition)
to cover the entire group
manifold, and second by giving a gauge
fixing (vector Drinfeld-Sokolov gauge)
which is convenient for arguing the (dis)connectedness
of the reduced phase space.
In sect.4 we move on to the actual
1 $+$ 1 dimensional case
for the WZNW $\rightarrow$ Liouville
reduction, where we shall see even more interesting
physical implications that are missing in the
0 $+$ 1 dimensional toy model.  We find, for instance,
that in the global
point of view, the singularities in
the classical Liouville solution are nothing but the points
where the Gauss decomposition breaks down,
and the number of those
points can be regarded as a conserved topological charge.
All of our analyses in this paper are purely classical.
Sect.5 will be devoted to our conclusions and outlooks.

\bigskip
\noindent
{\bf 2. $SL(2,\R)$
Toda (Liouville) theory in 0 $+$ 1 dimension}
\medskip

For a simple illustration,
we shall begin with a toy model where the
aspects of being global appear dramatically.
We first define our 0 $+$ 1 dimensional
model in the Hamiltonian formalism,
and then provide
a heuristic argument signifying
the global aspects in the Lagrangian
formalism.
Later we return
to the Hamiltonian formalism to examine the model in
more detail.

\medskip
\noindent
{\sl 2.1. Hamiltonian reduction}

Consider a
point particle moving on the group manifold $G$.
The description of the model as a Hamiltonian system
is standard:
the phase space is the cotangent bundle of
the simple Lie group $G$,
$$
M = {\rm T}^*G
  \simeq \{(g,J)\, \vert\,\, g \in G, \, J \in {\cal G}\} ,
\eqno(2.1)
$$
where the fundamental Poisson brackets
are $\bigl\{ g_{ij}\, , g_{kl} \bigr\} = 0$ and
$$
\bigl\{ g_{ij}\, , \tr(T^aJ) \bigr\} = (T^a g)_{ij}, \quad
\bigl\{ \tr(T^aJ)\, , \tr (T^bJ)\bigr\} = \tr ([T^a, T^b] J),
\eqno(2.2)
$$
with $T^a$ being a basis set of matrices in
some irreducible representation of ${\cal G}$.
The Hamiltonian is then
$$
H = {1\over2} \tr J^2,
\eqno(2.3)
$$
which yields the dynamics,
$$
\dot g = \{ g \, , H\} = Jg, \qquad \dot J = \{ J\, , H\} = 0.
\eqno(2.4)
$$
Hence the particle (whose position is given by the value
$g(t)$ on $G$) follows the free motion on the
group manifold,
${d\over{dt}}(\dot g g^{-1}) = 0$.
Incidentally, we note that
the \lq right-current',
$$
\tilde J := - g^{-1} J g,
\eqno(2.5)
$$
commutes with the \lq left-current'
$J$ and forms the Poisson brackets analogous to (2.2),
$$
\bigl\{ g_{ij}\, , \tr(T^a \tilde J) \bigr\}
= - (g T^a)_{ij}, \quad
\bigl\{ \tr(T^a \tilde J)\, , \tr (T^b \tilde J)\bigr\}
= \tr ([T^a, T^b] \tilde J),
\eqno(2.6)
$$

Now let us take $G = SL(2,\R)$ with its defining
representation for (2.2).
The constraints we will impose are
precisely the same as those imposed
in the WZNW $\rightarrow$ Liouville reduction [4],
$$
\tr(e_{12}J) = \mu \qquad {\rm and}
\qquad \tr(e_{21}\tilde J) = -
\nu,
\eqno(2.7)
$$
where
$\mu$ and $\nu$ are constants, and
$e_{ij}$ are the usual matrices having 1 for
the $(i,j)$-entry and 0 elsewhere.
It is almost trivial to see that
the two constraints in (2.7) are first class.

To be more explicit, let us parametrize the phase space
$M$ as
$$
g = \left(\matrix{
g_{11} & g_{12} \cr
g_{21} & g_{22} \cr}
\right),
\qquad
J = \left(\matrix{
j_0 & j_+ \cr
j_- & -j_0 \cr}
\right).
\eqno(2.8)
$$
Then
the first constraint in (2.7)
is just $j_- = \mu$ while the second implies
$$
\mu g_{12}^2 - 2j_0\,g_{22}g_{12} - j_+ g_{22}^2 = - \nu.
\eqno(2.9)
$$
{}From this we see that on the hypersurface
$g_{22} = 0$ in the constrained submanifold $\mc \subset M$
defined by (2.7) we must have $g_{12} = \sqrt {- \nu/\mu}$ if
$\mu\nu < 0$.  But if $\mu\nu > 0$,
then there is no solution for
$g_{12}$ on the hypersurface $g_{22} = 0$ in $\mc$,
which implies that $M_c$, and consequently also
the reduced phase space $\mr$,
is {\it disconnected} ---
one with $g_{22} > 0$ and the other
with $g_{22} < 0$.  Recall that the Gauss
decomposable patch is of the form (1.3), which for
$SL(2,\R)$ reads
$$
g = g_+\cdot g_0 \cdot g_-
  =
\left(\matrix{
            1           &         a      \cr
            0           &         1      \cr}\right)
    \left(\matrix{
          e^{{x\over2}} &         0      \cr
            0           & e^{-{x\over2}} \cr}\right)
    \left(\matrix{
            1           &         0      \cr
            c           &         1      \cr}\right)\ .
\eqno(2.10)
$$
It is then easy to see that on
the Gauss decomposable patch we always have
$g_{22} > 0$.
Hence, when $\mu\nu > 0$
the conventional description of the
reduced theory is actually sufficient (self-contained)
as a description
of the subsystem on the Gauss decomposable
patch, as it is anyway decoupled from the subsystem
for which $g_{22} < 0$.
We shall discuss this case $\mu\nu>0$
more generally in sect.3.
In the rest of this section, however,
we shall concentrate on
the case $\mu\nu < 0$, where the importance of
the global consideration is more transparent.
(Below we set $\mu = - \nu = 1$ for simplicity.)

\medskip
\noindent
{\sl 2.2. Lagrangian description ---
what is the fate of the particle?}

We now illustrate --- through a heuristic
argument --- an interesting global aspect of the
reduced system.
Consider first
the Gauss decomposable patch (2.10).
Observe that the local gauge
transformations
generated by the constraints (2.7) are
$$
g \longrightarrow \alpha\,g\,\gamma, \qquad
J \longrightarrow \alpha\,J\,\alpha^{-1},
\eqno(2.11)
$$
where $\alpha = e^{\theta e_{12}}$ and $\gamma = e^{\xi e_{21}}$.
Thus in the Lagrangian formalism, the present constrained
system is realized as a gauge theory [4,5] possessing
the local symmetry
under (2.11), with the help of Lagrange multipliers.
Choosing the  `physical gauge'  $a = c = 0$,
one can eliminate the Lagrange multipliers
using their equations of motion.
This yields the effective, reduced Lagrangian,
$$
L_{\rm red} = {1\over4}\dot x^2 + e^x\ .
\eqno(2.12)
$$
Thus, as long as we are on the Gauss decomposable
patch, we obtain as the reduced subsystem
a particle moving on a line
under the influence of the exponential potential
$V_{\rm red} = -e^x$.
The equation of motion derived from $L_{\rm red}$ is hence
$$
\ddot x = 2 e^x,
\eqno(2.13)
$$
which has, for \lq energy' $E := {1\over4}\dot x^2 - e^x < 0$,
the general solution
$$
x(t) = - 2 \ln \Bigl({{\cos{\omega(t - t_0)}}\over\omega}\Bigr)\ ,
\eqno(2.14)
$$
where
$\omega = \sqrt{\vert E \vert}$ and
$t_0$ are constants
determined from the initial condition given.
For instance, for the initial condition
$x(0) = 0$ and $\dot x(0) = 0$, we have $\omega = 1$, $t_0 = 0$.
We then observe that the particle reaches the infinity $ x = \infty$
with the finite time $ {\pi\over2}$.
Hence, if this patch with $-\infty < x < \infty$ was the
only \lq world' for the particle, then the particle would
sooner or later face a \lq catastrophe', as long as the energy
is negative.
But fortunately, we know that there
are
other \lq worlds' (patches)
in the group manifold where the particle could live
after it leaves the original patch.
But then, what happens to the particle
after it experiences the catastrophe?

To know this we
first recall [4] that the entire $SL(2,\R)$
group manifold can be covered by
the four patches:
$$
g = g_+\cdot g_0 \cdot g_- \cdot \tau \qquad
{\rm with}
\qquad
\tau = \pm 1, \quad \pm
\left(\matrix{
0  & 1 \cr
-1 & 0 \cr}
\right).
\eqno(2.15)
$$
It is then straightforward to see (by proceeding similarly
as before) that,
on the two patches $\tau = \pm 1$,
$g_{22}$ never vanishes
($g_{22} > 0$ for $\tau = 1$ and
$g_{22} < 0$ for $\tau = -1$) and
moreover the reduced Lagrangian $L_{\rm red}$
takes the same form (2.11).
Thus the hypersurface $g_{22} = 0$ is exactly the place where the local
description using the two patches breaks down.
(The other two patches, where $g_{22}$ can become zero,
admit neither a
convenient gauge like the physical gauge nor a regular description at
$g_{22} = 0$, as we will discuss shortly.)
But since
$g_{22} = 0$ is merely a lower dimensional submanifold
in the entire
group manifold, we shall for the moment disregard the
singular hypersurface (the \lq domain-wall' between the two good
\lq domains' $\tau = \pm 1$) and
consider the motion of the particle only on the
patches $\tau = \pm 1$.

We notice at this point that
since the values of $x$ in the two patches
are defined seperately on each patch and also depend
on the gauge fixing condition, we
must provide a method to extract the \lq physical position' of the
particle which has gauge- and patch-independent meaning.
This may be accomplished if we
identify those $g$ which are gauge equivalent under
(2.11); namely,
we define the physical position of the particle by the values of
$g \in SL(2,R)/GL(1)_{\rm left}\times GL(1)_{\rm right}$.
For instance, for $\lambda \ne 0$,
the two points,
$$
\left(\matrix{
\lambda  &    0        \cr
 0      & \lambda^{-1} \cr}
\right)
\qquad
{\rm and}
\qquad
\left(\matrix{
 0      &    1        \cr
 -1     & \lambda^{-1} \cr}
\right)
\eqno(2.16)
$$
are gauge equivalent and hence may be regarded identical.
Since in the patch $\tau = 1$ the physical gauge corresponds to
the first one in (2.16) with
$\lambda = e^{{x}\over2}$, in the second gauge in (2.16)
the trajectory of the particle
moving from $x = 0$ to $\infty$ may be represented
symbolically as
$$
\left(\matrix{
 0      &    1        \cr
 -1     &    1        \cr}
\right)
\quad
\longrightarrow
\quad
\left(\matrix{
 0      &    1        \cr
 -1     &    {1\over\infty}  \cr}
\right).
\eqno(2.17)
$$
In the patch $\tau = -1$, on the other hand, the physical gauge
corresponds to $\lambda = - e^{x\over2}$ and
one can show by an analogous but \lq converse' argument that
when a particle appears at the catastrophic point $x = \infty$
it would take the converse
path to the above and reaches the point $x = 0$ in the
patch if it
has a sufficient energy.  In the new gauge this passage
reads
$$
\left(\matrix{
 0      &    1        \cr
 -1     &    - {1\over\infty}  \cr}
\right)
\quad
\longrightarrow
\quad
\left(\matrix{
 0      &    1        \cr
 -1     &   - 1        \cr}
\right).
\eqno(2.18)
$$
But the fact that the final point
in (2.17) and the initial point in (2.18) are identcal
(although that point does not belong to
the two patches) implies that
the particle
{\it oscillates} with the period $2\pi$
between the two points $x = 0$ in the two
patches.

Clearly, what we need is a description of the redecued
theory valid
globally even on a patch which contains the catastrophic
point.  In fact, one can derive the reduced Lagrangian
for any of the other two
patches by choosing a gauge fixing condition properly.  By doing
this one finds that the reduced Lagrangian always takes the form
(2.12) but it contains a singularity at the
hypersurface $g_{22} = 0$.  (Since
$g_{22}$ is gauge invariant, this statement
is gauge independent.)  This singularity stems from the fact that
in the original configuration space $ G $
the gauge group $GL(1)_{\rm left}\times GL(1)_{\rm
right}$ does not act freely on the hypersurface
$g_{22} = 0$ and, as a result,
the Lagrangian description of the reduced theory
necessarily suffers from the singularity.
By contrast,
the gauge group does indeed act freely in
the phase space $M = {\rm T}^*G$ even on the hypersurface
$g_{22} = 0$,
it is therefore possible to have a globally
well-defined description of the reduced theory using
the Hamiltonian formalism.  Next, we wish to find it out
explicitly.

\medskip
\noindent
{\sl 2.3.  Oscillation in the Hamiltonian description}

For a description of the reduced system in the Hamiltonian
formalism,
there can be two options; one by the Dirac approach (by
introducing a gauge fixing condition
and the Dirac brackets), and the
other by the \lq gauge invariant approach'.  The latter
begins with choosing a set of
gauge invariant functions on $\mc$
for a set of coordinates of the
reduced phase space $\mr$, adopting the original Poisson brackets
(2.1) as a basis for computing the reduced Poisson brackets.
Since at the moment we do not have a
convenient,
global gauge fixing free from singularity we will pursue
the latter approach.

Finding a gauge invariant basis set of functions on $\mc$ would be
easier if we could use a \lq minimum'
coordinate system of $\mr$ which is, of
course,
of dim$\,\mr = 2$.
In the present case this could be achieved by eliminating four
variables by solving (2.9) after setting $j_- = 1$ and using
two gauge fixing conditions.
This however leads to either non-polynomial
expressions for gauge invariant functions or
a singularity in $g_{22}$ again.  We shall circumvent this
by not solving (2.9) explicitly but allowing an extra
variable for the coordinate of $\mr$ keeping (2.9) in
mind.  In this spirit we shall find a set of
three gauge invariant functions on $\mc$ having a relation among
them so that dim$\, \mr = 2$.

One can find easily such a set
of functions which are invariant
under the gauge transformations (2.11):
$$
Q = g_{22},
\qquad P = g_{12} - j_0 g_{22},
\qquad H = j_0^2 + j_+.
\eqno(2.19)
$$
Here $H$
is the Hamiltonian (2.3) on $\mc$ whereas $P = \dot Q = \{Q\, , H\}$.
The set $(Q,H)$ is actually enough to serve as a basis for gauge
invariant functions $f(g,J)$ on $\mc$, because
there exists a gauge fixing procedure which yields the two
invariants ({\it i.e.}, the gauge in which
$j_0 \rightarrow 0$, $\tilde j_0
\rightarrow 0$ with $\tilde j_0$ defined from $\tilde J$
analogously to $j_0$ in (2.8)).
However, the above set does not
form a \lq polynomial basis' ({\it i.e.} in terms of
which any polynomial
gauge invariant function can be expressed polynomially), for $P$
cannot be expressed polynomially in terms of them.
In fact, the relation between the three invariants is
$$
H Q^2 = P^2 - 1.
\eqno(2.20)
$$
Thus we shall regard
the two dimensional surface determined by
(2.20) in the three dimensional space $(Q, P, H)$
as the reduced phase space $\mr$.
The variables of the space form a
polynomial closed algebra under the Poisson brackets:
$$
\{Q\, , P\} = {{Q^2}\over2}, \qquad
\{Q\, , H\} = P, \qquad
\{P\, , H\} = Q H.
\eqno(2.21)
$$
Combining the last two equations in
(2.21) and the fact that the
Hamiltonian $H$ is a constant of motion we find,
for $H = E = - \omega^2 < 0$, that $Q$ obeys
the equation for a harmonic oscillator,
$$
\ddot Q + \omega^2 Q = 0.
\eqno(2.22)
$$
This result is consistent with (2.13) and (2.14)
on account of (2.19) and (2.20).
The oscillatory motion of the particle can also be seen in
the phase space $\mr$ if we slice the surface by a
(negative) constant $H$, as it forms the ellipse,
$P^2 + \omega^2 Q^2 = 1$.

We now provide a set of Hamiltonian subsystems
to comprise the reduced Hamiltonian system on $\mr$.
In terms of manifolds $M_k$, Poisson brackets $\{\, ,\}_k$ and
Hamiltonians $H_k$ for $k = 1, 2, 3, 4$,
the first and the third subsystems are given as
$$
\eqalignno{
&M_1 = \{\, (Q, P)\, \vert\, Q > 0,
\quad -\infty < P < \infty \, \},
\cr
&\{Q\, , P\}_1 = {{Q^2}\over2},
\qquad
H_1(Q,P)  = {1\over{Q^2}}(P^2 - 1),
&(2.23)
}
$$
and
$$
\eqalignno{
&M_3 =
\{\, (Q, P)\, \vert\, Q < 0, \quad -\infty < P < \infty \, \},
\cr
&\{Q\, , P\}_3 = {{Q^2}\over2},
\qquad
H_3(Q,P)  = {1\over{Q^2}}(P^2 - 1),
&(2.24)
}
$$
while the second and the fourth are
$$
\eqalignno{
&M_2 = \{\, (Q, H)\, \vert\,  -\infty < Q < \infty,
         \quad H Q^2 + 1 > 0 \, \},
\cr
&\{Q\, , H\}_2 = \sqrt{ H Q^2 + 1 },
\qquad
H_2(Q,H)  = H,
&(2.25)
}
$$
and
$$
\eqalignno{
&M_4 = \{\, (Q, H)\, \vert\,  -\infty < Q < \infty,
         \quad H Q^2 + 1 > 0 \, \},
\cr
&\{Q\, , H\}_4 = - \sqrt{ H Q^2 + 1 },
\qquad
H_4(Q,H)  = H.
&(2.26)
}
$$
When these subsystems are glued together
they constitute the reduced Hamiltonian system.
The transition
between the subsystems is done through the relation (2.20).
In effect, the subsystem $M_1$  (resp.~$M_3$) represents the $Q>0$
(resp.~$Q<0$) part of the surface (2.20), and
the subsystem $M_2$ (resp.~$M_4$) represents the $P>0$ (resp.~$P<0$)
part, respectively.
Thus we conclude that the reduced Hamiltonian system is perfectly
well-defined in terms of the four local Hamiltonian
subsystems.

\bigskip
\noindent
{\bf 3. $SL(n,\R)$ Toda theory in 0 $+$ 1 dimension}
\medskip

In sect.2 we have seen by the 0 $+$ 1
dimensional toy model that the global viewpoint can change
the interpretation of the dynamics drastically.
In this section we give a general framework for
$G = SL(n,\R)$ to deal with
the global reduced system, by generalizing
the idea used in the previous section.

\medskip
\noindent
{\sl 3.1. Global reduction --- the Bruhat decomposition}

Analogous to the previous $SL(2,\R)$ case, the WZNW reduction
to Toda theories is defined to the Hamiltonian system
(2.1) -- (2.3) by imposing a set of first
class constraints.  For $G = SL(n,\R)$ these constraints
are defined [2] by generalizing (2.7) as
$$
\pi_- (J)=I_-
\qquad\hbox{and}\qquad
\pi_+ (\tilde J )=- I_+,
\eqno(3.1)
$$
where
$$
I_-=\sum_{\alpha\in \Delta} \mu_\alpha E_{-\alpha},
\quad
I_+=\sum_{\alpha\in \Delta} \nu_\alpha E_{\alpha}
\qquad
\hbox{with}\quad \mu_\alpha\neq 0,\quad \nu_\alpha\neq 0\ .
\eqno(3.2)
$$
Here $\Delta$ is the set of simple roots, the
$\mu_\alpha$, $\nu_\alpha$ are {\it nonzero} constants
associated to the step generators $E_{\mp \alpha}$,
and the projections
$\pi_\pm$ in (3.1) refer to the subalgebras
${\cal G}_\pm$ mentioned in sect.1.
As before, these constraints generate a gauge
symmetry of the type (2.11) with $\alpha \in e^{{\cal G}_+}$
and $\gamma \in e^{{\cal G}_-}$.

Since the obstacle for
a global description is the intrinsic locality of
the Gauss decomposition
(1.3), it is natural to seek for a
set of patches that cover the entire group manifold $G$
having
the Gauss decomposable patch in it.
A natural choice is given by
the {\it Bruhat} (or {\it Gelfand-Naimark})
 {\it decomposition} [6],
$$
g_m = g_+ \cdot m \cdot g_0 \cdot g_- \ ,
\eqno(3.3)
$$
where $m$ is a diagonal matrix given, for $G = SL(n,\R)$, by
$$
m={\rm diag}\left(m_1,m_2,\ldots,m_{n}\right),
\qquad \hbox{with} \quad
m_i=\pm 1,\quad \prod_i m_i =1.
\eqno(3.4)
$$
Obviously, there are $2^{n-1}$ possibilities for $m$.
With these $m$ the entire group
manifold $G$ can be decomposed as
$$
G=\bigcup_m G_m \bigcup G_{\rm low},
\eqno(3.5)
$$
where $G_m$ corresponds to the `domain' labelled by
$m$ whereas $G_{\rm low}$ is a union of
\lq domain-walls', {\it i.e.,} certain lower dimensional
submanifolds of $G$.  We note that
these domains are
disjoint, and
the decomposition (3.3) of every $g\in G_m$ is unique.
It is also worth noting that
in this $G=Sl(n,{\bf R})$ case the
union of the domains $\bigcup_m G_m$
is the open submanifold
consisting of matrices with nonzero
prinicipal minors whose signs are specified by $m$,
which is possible in $2^{n-1}$ different ways
for an $n\times n$ matrix of determinant 1 and nonzero minors.
Correspondingly, $G_{\rm low}$
consists of matrices with unit determinant and
at least $1$ vanishing principal minor.

Because of the factor $m$ in the decomposition (3.3),
we obtain in general the reduced dynamics slightly
different from the familiar Toda dynamics.
More precisely, one can derive
(by the conventional reduction
procedure where one chooses the physical gauge) the
reduced Lagrangian governing the dynamics
on the patch $G_m$,
$$
L_m
= {1 \over 2}\, \tr\,
\bigl(g_0^{-1}\dot g_0\bigr)^2
- \tr\, \bigl(I_- g_0\, m\, I_+ \,
m^{-1} g_0^{-1}\bigr)\ .
\eqno(3.6)
$$
For $m = \pm 1$,
one sees
upon using (1.5) and
$K_{ij} = {{\vert \alpha_i \vert^2}\over 2}\tr (H_i H_j)$
that the Lagrangian (3.6) reduces to the
standard Toda Lagrangian (1.1)
in 0 $+$ 1 dimension; otherwise
it differs from the standard one in general.
Thus the reduced system
consists of many \lq quasi-Toda' subsystems, among which
the standard Toda appears
on the trivial patch $m = 1$.  Symbolically,
we may therefore write the actual reduction as
$$
{\rm WZNW \longrightarrow Toda \oplus (Toda)' \oplus
(Toda)''\oplus \cdots\ .}
\eqno(3.7)
$$
We are now interested in the
question whether these different subsystems
are disconnected or not.

\medskip
\noindent
{\sl 3.2. Vector Drinfeld-Sokolov gauge}

In sect.2 we have seen, based on the consistency of the
constraint (2.8) against the condition $g_{22} = 0$,
that the
reduced phase space $M_{\rm red}$ is disconnected
if $\mu\nu > 0$.
Now we shall try to examine the (dis)connectedness
generally for $SL(n,\R)$.  A convenient method for
doing this is to employ, instead of the physical gauge,
the so-called
Drinfeld-Sokolov
(DS) gauge [7] {\it both} for the left and right currents.
(The DS gauge has been used to
obtain a ${\cal W}$-algebra basis in the
WZNW (Kac-Moody) reduction [2].)  For simplicity,
in the rest of this section we just consider
the case where all the $\mu$'s and $\nu$'s are positive
and set all of them to unity.

We call the gauge {\it vector DS gauge} for $SL(n,\R)$
if the left and right currents are in the form,
$$
J = I_- + \sum_{i = 2}^n u_i e_{1i}
\qquad
\hbox{and}
\qquad
\tilde J = - I_+ - \sum_{i = 2}^n u_i e_{i1}.
\eqno(3.8)
$$
It is clear that since the usual, \lq chiral'
DS gauge is well-defined
({\it i.e.}, it is attainable and specifies
a unique representative among
the gauge equivalent points in the
phase space) for a chiral Kac-Moody current
under the same setting of gauge group,
the above vector DS gauge
is also well-defined.
Now, it follows from (2.5) and
the second equation in (3.8) that
the matrix $g$ has the same value along each
anti-diagonal line:
$$
g_{ij} = g_{kl}, \qquad \hbox{if} \quad i + j = k + l.
\eqno(3.9)
$$
It also follows that
$$
g_{1,j-1} = \sum_{i = 2}^n u_i g_{ij}, \qquad \hbox{for}
    \quad j = 2, \ldots, n\ ,
\eqno(3.10)
$$
which shows that
all the \lq lower' entries
$g_{1,j-1}$ in the first row for $1 \leq j-1 \leq n-1$
can be expressed in terms of the
\lq higher' ones together with $u_i$'s for $2 \leq i \leq n$.
Thus, if we set
$$
u_{i+j} := g_{ij},  \qquad \hbox{for} \quad
  n+1 \leq i + j \leq 2n,
\eqno(3.11)
$$
then, combinging with the first equation of (3.8),
the reduced phase space $\mr$
is parametrized by the set of variables,
$u_2$, $u_3, \ldots, u_{2n}$,
which are subject
to the condition $\det g = 1$.  Hence the total
dimension of the reduced phase space is
dim$\, M_{\rm red} = (2n-1) - 1 = 2(n-1)$.
Accordingly,
in the vector DS gauge
the matrix $g$ can be written as
$$
g = \pmatrix{
\,g_{11}(u)\hfill & g_{12}(u)\hfill & \ldots
                  & g_{1,n-1}(u)\hfill & u_{n+1}\hfill \cr
\,g_{12}(u)\hfill & g_{13}(u)\hfill & \ldots
                  & u_{n+1}\hfill      & u_{n+2}\hfill \cr
\,\, \vdots\hfill &
     \, \vdots\hfill &&
        \, \vdots\hfill &
           \, \vdots\hfill &\cr
\,g_{1,n-1}(u)\hfill & u_{n+1}\hfill & \ldots
                    & u_{2n-2}\hfill     & u_{2n-1}\hfill \cr
\,u_{n+1}\hfill      & u_{n+2}\hfill & \ldots
                    & u_{2n-1}\hfill     & u_{2n}\hfill   \cr
}\ .
\eqno(3.12)
$$


However, the objects of our concern are not
the $u_i$'s now, but their gauge invariant functions
defined by the principal minors of the matrix $g$,
$$
Q_n := u_{2n},
\qquad
Q_{n-1} := \det\pmatrix{u_{2n-2} & u_{2n-1}   \cr
                        u_{2n-1} & u_{2n}   \cr},
\qquad
Q_{n-2} := \cdots,
\eqno(3.13)
$$
for $Q_i$ with $i = 2$, $3$, $\ldots$, $n$, that is,
those $n-1$ principal minors constructed from the
lower right corner of the matrix $g$.
It is not difficult to see that they are gauge
invariant under (2.11).  On the
Gauss decomposable patch the Toda variables
$\varphi^i$ in (1.5) are directly related to those
principal minors by $e^{-\varphi^i/2} = Q_{i+1}$,
but unlike the Toda variables
these $Q_i$ are globally well-defined over the
entire reduced phase space.  These $Q_i$ are a generalization
of the $Q$ variable used in the previous section.

\medskip
\noindent
{\sl 3.3. Is Heaven connected with Hell?}

As in the case of $SL(2,\R)$, we want to know the global
structure of the reduced phase space in the $SL(n,\R)$
case.  Although the entire global structure
for a generic $n$ seems hard to know, we can at least ease
the problem by restricting ourselves to the
simpler question whether
the standard Toda theory, that is, the reduced
subsystem on the Gauss decomposable
patch, is disconnected from the rest of the subsystems or not.

More precisely,
we shall ask whether there exists a smooth path
connecting the domain,
$$
Q_2 > 0, \quad Q_3 > 0, \quad \cdots, \quad  Q_n > 0,
\eqno(3.14)
$$
and the remaining domains
of the reduced phase space.  (We note that for a manifold
\lq connectedness' and `path-connectedness' are the same.)
We call the domain (3.14) --- which is the domain for the
standard Toda subsytem --- simply \lq Heaven',
and the remainder \lq Hell'.
Suppose now that there exists such a path connecting
Heaven and Hell.
Let $t$ be a real parameter
of the path, and $t_0$ the time
passing the border between Heaven and Hell,
$$
Q_2(t_0) \, Q_3(t_0) \cdots Q_n(t_0) = 0.
\eqno(3.15)
$$
The path is assumed to enter into Heaven immediately
after $t_0$, {\it i.e.},
at $t = t_1 = t_0 + \epsilon$
with any infinitesimal $\epsilon$, we have
$$
Q_2(t_1) > 0,\quad Q_3(t_1) > 0, \quad \cdots, \quad
Q_n(t_1) > 0.
\eqno(3.16)
$$
In the following we argue that the answer to the question
is negative, that is, there exists no such path
for some $n$.

For this purpose it is useful to consider the two steps,
described below.  Calling
the point
$$
Q_2(t_0) = Q_3(t_0) = \cdots = Q_n(t_0) = 0\ ,
\eqno(3.17)
$$
\lq gate', we wish to argue along the line of
the following two statements:

\item{(i)} {\it There exists no
path entering
from Hell to Heaven through the gate.}
\item{(ii)} {\it Any path which
enters into Heaven
from Hell
must pass the gate.}

\noindent
Let us prove (i) for $n = 2$ mod 4, and $n = 3$.
For this, we observe first that if (3.17) holds then
$0 = Q_n(t_0) = u_{2n}(t_0)$
and
$0 = Q_{n-1}(t_0) = - u_{2n-1}^2(t_0)$, that is,
we get $u_{2n}(t_0) = u_{2n-1}(t_0) = 0$.  Repeating
this process we find that (3.17) actually means
$$
u_{n+2}(t_0) = u_{n+3}(t_0) = \cdots = u_{2n}(t_0) = 0,
\eqno(3.18)
$$
{\it i.e.}, all the entries of $g$ lower than the anti-diagonal
line are zero.
Hence, evaluating the determinant of $g$ at $t = t_0$ we get
$$
1 = \det g(t_0) = (-1)^{P[n]} u_{n+1}^n(t_0),
\eqno(3.19)
$$
where $P[n] := n(n-1)/2$ is the factor of permutation attached.
This result (3.19) shows that, for $n = 2$
mod 4, there is no real solution for $u_{n+1}(t_0)$, that is,
the gate point (3.17) does not even exist in the reduced phase
space $M_{\rm red}$.  This concludes the proof for
$n = 2$ mod 4.
Note that the gate point does exist
in $\mr$ in other cases with the solutions
$$
u_{n+1}(t_0) = \cases{ \pm 1, & if $n = 0$ mod 4; \cr
                          +1, & if $n = 1$ mod 4; \cr
                          -1, & if $n = 3$ mod 4. \cr}
\eqno(3.20)
$$
Thus, to argue for other cases we need something more.

Now we consider the case $n = 3$.
{}From (3.16) we must have
$$
Q_3(t_1) = u_6(t_1) > 0\ , \qquad
Q_2(t_1) = u_4(t_1) u_6(t_1) - u_5^2(t_1) > 0\,.
\eqno(3.21)
$$
On the other hand, on account of
the smoothness assumption we made for
the path, the condition
$u_4(t_0) = -1$ in (3.20) implies
$u_4(t_1) < 0 $, which contradicts
with (3.21) above.  We therefore have shown the statement
(i) for $n = 3$.  One can prove also for $n = 4$ using a
slightly
involved but similar argument, for which we  refer to [8].
The statement (i) has not been
(dis)proven for a generic $n$, yet.

Since the second statement (ii) seems even harder to
argue in general,
we prove here for
$n = 3$ only (the case $n = 2$ is trivial),
and again for $n = 4$ we refer to [8].
Suppose that there exists a path entering from Hell
to Heaven without passing the gate.  The possibilities are
thus either
$$
Q_3(t_0)=0
\quad\hbox{and}\quad
Q_2(t_0)>0
\eqno(3.22)
$$
or
$$
Q_3(t_0)>0
\quad\hbox{and}\quad
Q_2(t_0)=0.
\eqno(3.23)
$$
The first possibility (3.22) can be denied at once since
it is inconsistent with the definitions for $Q_2$ and
$Q_3$ (cf.(3.21)).
To deny (3.23), we just use
$0 = Q_2(t_0) = u_4(t_0) u_6(t_0) - u_4^2(t_0)$
to obtain
$$
\eqalignno{
0 < Q_3^3(t_0) \det g(t_0) &=
    u_6^3(t_0) \det
   \pmatrix{ g_{11}  &   g_{12} &  u_4 \cr
             g_{12}  &      u_4 &  u_5 \cr
              u_4    &      u_5 &  u_6 \cr} (t_0)  \cr
    &= - \bigl\{ u_5^3(t_0) - g_{12}(t_0)
                    u_6^2(t_0)\bigr\}^2 \ ,
&(3.24)
}
$$
which is, again,
a contradiction.
Thus, combining the result obtained
earlier, we have learned that for $n = 2$, $3$ and 4,
the Toda subsystem
is disconnected from all the rest of the subsystems
in the reduced system.

\bigskip
\noindent
{\bf 4. $SL(2,\R)$ Toda (Liouville) theory in 1 $+$ 1
dimensions }
\medskip

Although it is not simple to discuss the generic $SL(n,\R)$
case, we can
proceed similarly for the 1 $+$ 1 dimensional
field theory case at least for $SL(2,\R)$, {\it i.e.}, for
the WZNW $\rightarrow$ Liouville reduction.
Our study of the global aspects in this
field theory case then reveals an intricate structure
that we did not find in the 0 $+$ 1 dimensional
counterpart in sect.2.  The relation to the singularity
in classical solutions will also be discussed at the end.

\medskip
\noindent
{\sl 4.1. Hamiltonian reduction}

We first recall the WZNW $\rightarrow$ Liouville
reduction starting, as in the 0 $+$ 1 dimensional case in sec.2,
with the Hamiltonian description of the WZNW model.
Consider the phase space of
the type (2.1) but with $(g(x^1), J(x^1))$ both periodic
functions in space $x^1$, and postulate the fundamental Poisson
brackets,
$$
\eqalign{
\{g_{ij}(x^1)\, , g_{kl}(y^1)\} &= 0, \cr
\{g_{ij}(x^1)\, , \tr(T^aJ(y^1)) \} &= (T^a g(x^1))_{ij}\,
       \d(x^1 -  y^1), \cr
\{\tr(T^aJ(x^1))\, ,
\tr (T^bJ(y^1))\} &= \tr ([T^a, T^b] J(x^1))\,
        \d(x^1 - y^1) + 2\, \tr(T^a T^b)\,\d'(x^1 - y^1),
}
\eqno(4.1)
$$
where $\d' = \pa_1 \d(x^1 - y^1)$.
The Hamiltonian is then taken to be
$$
H = \int dx^1 \, {1\over4} \tr (J^2 + \tilde J^2),
\eqno(4.2)
$$
which yields the field equations
$$
\dot g = \{ g\, , H \} = J g - g^\prime, \qquad
\dot J = \{ J\, , H \} = J^\prime,
\eqno(4.3)
$$
or $\pa_-(\pa_+ g\, g^{-1}) = 0$.
The right-current, which acts as the generator for the
right-transformation, reads
$$
\tilde J = - g^{-1} J g + 2 g' g^{-1}.
\eqno(4.4)
$$
Upon imposing the constraints (2.7), which
are still first class under the Poisson brackets (4.1),
we have the gauge transformations generated by them,
$$
g \longrightarrow \alpha\,g\,\gamma, \qquad
J \longrightarrow \alpha\,J\,\alpha^{-1} + 2 \alpha' \alpha^{-1}.
\eqno(4.5)
$$

In chiral currents the constrains (2.7) are formally
the same as before, but
in components they are not quite so ---
the first constraint in (2.7)
is unchanged $j_- = \mu$ whereas in view of (4.4)
the second one now takes the form,
$$
\mu g_{12}^2 - 2j_0\,g_{22}g_{12} - j_+ g_{22}^2
+ 2 (g_{22}g_{12}' - g_{12} g_{22}') = - \nu.
\eqno(4.6)
$$
At first sight it appears that
the additional term in (4.6) does not alter the
condition for disconnectedness of the reduced phase space observed
in the toy model.  However, due to the space dimension $x^1$ the
previous argument must be modified.
In fact, we cannot rule out now the possibility of
having $g_{22}(x^1) = 0$ with $g_{22}'(x^1) \ne 0$ at {\it
some points} of $x^1$ (not all $x^1$),
in which case the above equation
has a
solution for $g_{12}(x^1)$.  In other words, the value of $g_{22}$
could go
out of the patches $\tau = \pm 1$
in some domain of $x^1$.
Thus we see that the description of the
Gauss decomposable patch
is not sufficient {\it irrespective} of the sign of $\mu\nu$.
An interesting observation is available at this point:
since for the case $\mu\nu > 0$
the configurations $g_{22} = 0$
{\it and} $g_{22}' = 0$ are incompatible at any point $x^1$,
one cannot shrink or extend
the loop of
$g_{22}$ --- the periodic condition in
$x^1$ implies that the configuration of $g_{22}$
can be expressed by a loop in $G$ ---
{\it across} the border
of the two patches $\tau = \pm 1$.
This shows that $g_{22}$ has a
conserved topological charge given by the number of zeros
$g_{22}(x^1)=0$ over the period in $x^1$.  The meaning of the
charge will be discussed later.
Hereafter, we shall consider only for the case $\mu\nu > 0$
(which is also the case of interest
from two dimensional gravity point of view)
and, for simplicity,
we shall set $\mu = \nu = 1$.

As in the toy model, we now try to give a global Hamiltonian
description in terms of local Hamiltonian subsystems.
Again, it is easy to find a set of gauge invariant differential
polynomials; one only has to change $H$ in (2.19) slightly,
$$
Q = g_{22},
\qquad P = g_{12} - j_0 g_{22},
\qquad V = j_0^2 + j_+ - 2j_0'.
\eqno(4.7)
$$
They form under the
Poisson brackets the following differential polynomial algebra:
$$
\eqalign{
\{ Q\, , Q \} & = 0, \cr
\{ P\, , P \} & = {1\over2}(Q^2)'\d + Q^2 \d', \cr
\{ P\, , V \} & = V Q \d + P \d' - Q\d'{}',
}
\qquad
\eqalign{
\{ Q\, , P \} & = {1\over2}Q^2 \d, \cr
\{ Q\, , V \} & = P \d - Q \d', \cr
\{ V\, , V \} & = 2 V' \d + 4 V \d' - 4 \d'{}'{}',
}
\eqno(4.8)
$$
where
$\{ Q\, , Q \} = \{ Q(x^1)\, , Q(y^1) \}$ and so on.
Note that besides the Virasoro subalgebra formed by $V$ there exists
another differential
polynomial subalgebra formed by $(Q, P)$.  Also, since we have
$$
\dot Q = \{ Q\, , H \} = P - Q',
\eqno(4.9)
$$
or $P = \pa_+Q$,
the fourth equation in (4.8)
shows that $Q$ is a conformal primary
field of weight $-{1\over2}$.
As in the toy model the three gauge invariant functions
are not independent but satisfy
$$
V Q^2 = P^2 - 2 Q' P + 2 Q P' + 1.
\eqno(4.10)
$$
This relation (which is in fact idential to (4.6))
may define the reduced phase space as a hypersurface
in the space spanned by $(Q(x^1), P(x^1), V(x^1))$.

In particular, for $Q \ne 0$
it is convenient to write
$$
Q = \pm e^{-{\p\over2}},
\eqno(4.11)
$$
and define the momentum conjugate to
$\p$ as
$\pi = {1\over Q}(Q' - P)$.
With these variables the Virasoro density $V$
takes the familiar form of the Liouville theory,
$$
\eqalignno{
V &= (\pi + {1\over2}\p')^2 - 2 (\pi + {1\over2}\p')' + e^{\p} \cr
  &= {1\over4}(\pa_+ \p)^2 - (\pa_+\p)' + e^{\p}.
&(4.12)
}
$$
One can derive the (global) field equation for $Q$ by
proceeding further from (4.9),
$$
\pa_+^2 Q = V Q    \qquad {\rm or} \qquad
\pa_+ Q \pa_-Q - Q\, \pa_+\pa_-Q + 1 = 0,
\eqno(4.13)
$$
where we have used (4.9), (4.10)
to obtain the second form of the field
equation.
Of course, for $Q \ne 0$ (4.13) reduces to
the Liouville equation,
$$
\pa_+\pa_-\p + 2 e^\p = 0,
\eqno(4.14)
$$
in the variable of (4.11).
In summary, we find that the reduced WZNW theory contains the
Liouville theory locally, and the reduced
WZNW theory consists of two copies of the Liouville theory
in the patches $\tau = \pm 1$ ({\it i.e.,} $Q > 0$ and $Q < 0$)
glued together by the domain-wall $Q = 0$.

\medskip
\noindent
{\sl 4.2. Global classical solution and the index of singularity}

{}From the above
construction and the relation between $Q$ and $\p$ we expect,
by using the globally defined
$Q$, that the singularities in the solution $\p$
of the Liouville theory disappear and moreover they
may be classified by means of
the topological charge carried by $Q$.
Let us examine these points through the solution
of the global equation (4.13) next.

The key observation for getting the solution of
the partial differential equation (4.13)
is the one we employed to solve the Liouville equation in the WZNW
context: it can be obtained from the WZNW solution, which is
trivial, by taking into account the constraints.  In our case,
since $Q$ is $g_{22}$ itself the WZNW solution
$g(x) = g^L(x^+)g^R(x^-)$ implies
$$
Q = g_{22} = g_{21}^L\,g_{12}^R + g_{22}^L\,g_{22}^R\ .
\eqno(4.15)
$$
{}From the fact that $J = \pa_+ g^L\,(g^L)^{-1}$ and
$\tilde J = - (g^R)^{-1}\pa_-g^R$ at on-shell
it follows that the constraints
(2.7) are equivalent to
$$
\eqalign{
&\pa_+ g_{21}^L\,g_{22}^L - \pa_+ g_{22}^L\,g_{21}^L = 1, \cr
&\pa_- g_{12}^R\,g_{22}^R - \pa_- g_{22}^R\,g_{12}^R = 1.
}
\eqno(4.16)
$$
This means that instead of the original second order partial
differential equation (4.13) we only have to deal with
the two linear ordinary
differential equations (4.16),
which are easy to solve.  Let us give a
quick way to reach the solution here.
The structure shared by both of the equations is $f'g - f g' = 1$.
For $fg \ne 0$ we
rewrite it as $(\ln {f\over g})' = {1 \over {fg}}$, which is
integrated to be
$$
{f\over g} = e^{\int {{dx}\over{fg}} + c } \equiv F(x),
\eqno(4.17)
$$
where $c$ is a constant of integration and
$F(x)$ is defined by this equation, which we regard as
an arbitrary function.
A differentiation of $F$ gives $F' = {1\over{g^2}}$, which shows
that we must have $F' > 0$.
The solution is then given by
$$
f = \pm {{F}\over {\sqrt{F'}}}, \qquad
g = \pm {{1}\over {\sqrt{F'}}}.
\eqno(4.18)
$$
Applying this procedure
to the equations (4.16), and combining with (4.15), we get
the solution for $Q$:
$$
Q(x) = \pm {{1 + F(x^+)G(x^-)}
\over{\sqrt{\pa_+F(x^+)\pa_-G(x^-)}}},
\eqno(4.19)
$$
where $F(x^+)$ and $G(x^-)$ are arbitrary functions of the argument
with $\pa_+F > 0$, $\pa_-G > 0$,
such that $Q$ be periodic in $x^1$.
For instance, if
we set the periodic condition $Q(x^0, x^1 +
2\pi) = Q(x^0, x^1)$, the choice,
$F(x^+) = \tan ux^+$, $G(x^-) = \tan vx^-$ with positive constants
$u$, $v$, satisfies the periodicity condition if
${1\over2}(u + v) = n$ is an (positive) integer.  Putting
${1\over2}(u - v) = r$ with $- n < r < n$
we find the solution
$$
Q(x) = \pm \cos (r x^0 + n x^1).
\eqno(4.20)
$$
On the other hand,
the global solution (4.19)
reduces on the patches $\tau = \pm 1$ to the local
one, namely the well-known Liouville solution,
$$
\p (x) = \ln \biggl(
   {{ \pa_+ F(x^+) \pa_- G(x^-)} \over {[1 + F(x^+)G(x^-)]^2}}
\biggr).
\eqno(4.21)
$$
The previous special solution (4.20) therefore leads to
$$
\p(x) = - 2 \ln \vert \cos (r x^0 + n x^1) \vert.
\eqno(4.22)
$$
We then notice that there appears in the
special solution essentially the same property
observed in the toy model ---
the apparently singular motion in the local
system is merely a part of the oscillation in the global system.
In general, the global dynamics of
physical quantities in the present model
could  turn out to be  regular
despite that the local counterpart might appear  singular.

Finally, let us detail the point of singularity slightly more.
{}From the above analysis we learn that
to any global (non-singular) solution $Q$ there
exists a local (possibly singular) Liouville solution $\p$
and that the converse is also true.
This implies that the singularity in the Liouville solution is
characterized by the zeros of $Q$.  More precisely, if we
define the singularity index $s$ of a Liouville solution by the
number of singular points at $x^0 =$ constant surface,
then $s$ is equal to the
number of zeros of $Q$ at
the fixed time.  But since the number of zeros of $Q$ is the
topological charge which is
{\it time-independent},
it can be used to classify the Liouville
solution.
For example, the previous special solution (4.20)
has index $s = 2n$
(in our case $s$ is always an even integer due to the periodic
condition).
On the other hand, since (4.15) is a linear differential
equation, one can in fact construct a
Liouville solution with any number of singular points $s$
by imposing
an appropriate initial condition for $Q$ with $s$ zeros.

\bigskip
\noindent
{\bf 5. Conclusions and outlooks}
\medskip

We have seen in this paper that
the reduced theory
obtained by the WZNW $\rightarrow$ Toda reduction
is not just the standard Toda theory but
has richer global structures which may bring
drastic changes in the interpretation of the dynamics.
More precisely, it was shown that
the reduced system actually
consists of many quasi-Toda subsystems where
the (dis)connectedness among them is crucial
in determining the global effects on the dynamics.
We also found that the 1 $+$ 1 dimensional `global Liouville
theory' constructed by the WZNW reduction possesses
a conserved topological charge which classifies
the classical solutions.
 Our general
framework for studying the global aspects of the
WZNW $\rightarrow$ Toda reduction uses the Bruhat
decomposition and the vector DS gauge, both of which
allow an immediate generalization to any other
simple Lie groups, not just to $SL(n,\R)$.

These results, although still preliminary, suggest
that those theories obtained by the WZNW reduction ---
whether or not they are in 0 $+$ 1 dimension or 1 $+$ 1
dimensions --- are interesting enough physically and
worth further investigation.  Some of the possible
directions are as follows:

\item{(1)} The analysis of
the (dis)connectedness of the subsystems in the reduced
theory is far from complete, since we know only for
$SL(n,\R)$ with $n = 2$, 3, 4 that the proper Toda subsystem is
disconnected from the rest.  It seems however reasonable
 to conjecture that
it is disconnected for any $n$.
Of course, we would like to have full information
on the global structure of the reduced theory, {\it e.g.},
as to how all those
subsystems are glued together for any possible
signs of $\mu_\alpha$, $\nu_\alpha$.
\item{(2)} More importantly, the change of the interpretation
in the nature of dynamics by being global could
imply a drastic change in the
theory at the quantum level as well.
An interesting question is whether it is possible to
construct a reasonable  quantum gravity model
 in two dimensions by quantizing
the 1 $+$ 1 dimensional
global Liouville theory obtained by the WZNW reduction
in the non-trivial topological sectors.
The difficulty is that in these sectors the energy is
not bounded from below in general [9],
similar to the case of the quasi-Toda subsystems (3.6)
of the toy model for generic $m$.
In the $SL(2,\R)$ toy model, a consistent
quantum mechanical version of the
reduced system has been recently constructed [10]
in the topologically non-trivial case $\mu\nu<0$.
\item{(3)}
Our classification  of the Liouville solutions by their
topological charge is related to earlier work [11] on the
singular sectors of the Liouville theory.
The dependence
of the topological charge on the `coadjoint orbit type'
of the
corresponding chiral Virasoro densities is analysed
in detail in [9], generalizing results of [12].
The precise relationships between these results as well as the
possible connections between the globally well defined
reduced WZNW systems and the ${\cal W}$-geometry theories
proposed in [13,14] as geometric reformulations of
Toda theories deserve further study.

We hope that the present paper has provoked the interest
of the reader.  More technical accounts of the global
aspects  of the reduced WZNW systems will appear elsewhere.

\vskip 1cm
\noindent
{\bf Acknowledgement:}
The authors wish to thank L.~O'Raifeartaigh, P.~Ruelle and
A.~Wipf for  useful discussions in the early stage of the work.
L.F. is also indebted to J.~Balog and L.~Palla for collaboration
on related matters.

\vfill\eject

\centerline{\bf References}

\vskip 0.8truecm

\item{[1]}
P. Bouwknegt and  K. Schoutens,
{\sl Phys. Rep.} {\bf 223} (1993) 183.

\item{[2]}
See, for example,
L. Feh\'er, L. O'Raifeartaigh, P. Ruelle, I. Tsutsui and A. Wipf,
{\sl Phys. Rep.} {\bf 222} (1992) 1.

\item{[3]}
E. Witten,
{\sl Commun. Math. Phys.} {\bf 92} (1984) 483.

\item{[4]}
P. Forg\'acs, A. Wipf, J. Balog, L. Feh\'er and L. O'Raifeartaigh,
{\sl Phys. Lett.} {\bf 227B} (1989) 214.

\item{[5]}
 J. Balog, L. Feh\'er, L. O'Raifeartaigh, P. Forg\'acs
and A. Wipf, {\sl Ann. Phys.} (N. Y.)  {\bf 203} (1990) 76.

\item{[6]}
See e.g.~Chapter IX in:
S. Helgason, ``Differential Geometry,
Lie Groups and Symmetric Spaces'',
Academic Press, New York, 1978.

\item{[7]}
V. G. Drinfeld and V. V. Sokolov, {\sl J. Sov. Math.} {\bf 30}
(1984) 1975.

\item{[8]}
L. Feh\'er and I. Tsutsui,
in preparation.

\item{[9]}
J. Balog, L. Feh\'er and L. Palla,  in preparation.

\item{[10]}
T. F\"ul\"op,
private communication.

\item{[11]}
A. K. Pogrebkov and M. K. Polivanov,
{\sl Sov. Sci. Rev. C. (Math. Phys.)} Vol.~5. (1985) pp. 197-272,
and references therein.

\item{[12]}
E. Aldrovandi, L. Bonora, V. Bonservizi and R. Paunov,
{\sl Int. J. Mod. Phys.} {\bf A9} (1994) 57.

\item{[13]}
J.-L. Gervais and Y. Matsuo,
{\sl Phys. Lett.} {\bf 274B} (1992) 309;
{\sl Commun. Math. Phys.}  {\bf 152} (1993) 317.

\item{[14]}
A. V. Razumov and M. V. Saveliev,
Differential geometry of Toda systems,
preprint LPTENS-93/45 (hep-th/9311167);
\item{}
J.-L. Gervais and M. V. Saveliev,
$W$-geometry of the Toda systems associated with non-exceptional
simple Lie algebras,
preprint LPTENS-93/47 (hep-th/9312040).

\bye